# Computational Modeling and Analysis of Diesel-fuel Injection and Autoignition at Transcritical Conditions

Matthias Ihme[1], Peter C. Ma[1] and Luis Bravo[2]

[1]Department of Mechanical Engineering, Stanford University, Stanford, CA 94305, USA.

E-mail:       mihme@stanford.edu
Telephone:    +(1) 650 724 3730
Fax:          +(1) 650 725 3525

[2]Propulsion Division, Vehicle Technology Directorate, U.S. Army Research Laboratory, Aberdeen Proving Ground, MD 21005, USA

E-mail:       luis.g.bravo2.civ@mail.mil
Telephone:    +(1) 410-278-9719
Fax:          +(1) 410-278-0590

**Abstract.** The need for improved engine efficiencies has motivated the development of high-pressure combustion systems, in which operating conditions achieve and exceed critical conditions. Associated with these conditions are strong variations in thermo-transport properties as the fluid undergoes phase transition, and two-stage ignition with low-temperature combustion. Accurately simulating these physical phenomena at real-fluid environments remains a challenge. By addressing this issue, a high-fidelity LES-modeling framework is developed to conduct simulations of transcritical fuel spray mixing and auto-ignition at high-pressure conditions. The simulation is based on a recently developed diffused interface method that solves the compressible multi-species conservation equations along with a Peng-Robinson state equation and real-fluid transport properties. LES analysis is performed for non-reacting and reacting spray conditions targeting the ECN Spray A configuration at chamber conditions with a pressure of 60 bar and temperatures between 900 K and 1200 K to investigate effects of the real-fluid environment and low-temperature chemistry. Comparisons with measurements in terms of global spray parameters (i.e., liquid and vapor penetration lengths) are shown to be in good agreement. Analysis of the mixture fraction distributions in the dispersed spray region demonstrates the accuracy in modelling the turbulent mixing behavior. Good agreement of the ignition delay time and the lift-off length is obtained from simulation results at different ambient temperature conditions and the formation of intermediate species is captured by the simulations, indicating that the presented numerical framework adequately reproduces the corresponding low- and high-temperature ignition processes under high-pressure conditions, whichare relevant to realistic diesel-fuel injection systems.

## 1. Introduction

As propulsion systems continue to push towards higher efficiencies, the demand for high-performance combustion devices operating at even-higher pressures has been steadily increasing. This trend concerns traditional applications, such as diesel and gas turbine engines, as well as rocket motors and pressure-gain combustion systems (Yang 2000; Chehroudi 2012). In these environments, reacting flows undergo exceedingly complex and interrelated thermophysical processes, starting with compressible fuel injection, atomization, mixing, and heating, leading to ignition and combustion. These processes are often subject to pressure levels well above the thermodynamic critical state. In the context of diesel engines, the compressed fluid is typically injected at subcritical temperatures into a mixture that undergoes a thermodynamic state transition into the supercritical regime. During this process, the fuel typically goes through the so-called pseudo-boiling region (Oschwald et al. 2006), which is depicted in Fig. 1 for *n*-dodecane. Manin et al. (2014) conducted high-speed long-distance microscopy measurements of *n*-dodecane sprays and showed that the interfacial behavior of fluids exhibits properties markedly different from classical two-phase breakup, including diminishing effects from surface tension and heat of vaporization. Dahms et al. (Dahms and Oefelein 2013; Dahms et al. 2016) presented a theoretical framework that explains the conditions under which a multi-component fluid mixture transitions from two-phase breakup to single-phase mixing in a manner consistent with experimental observations. The theory is based on coupled real-fluid thermodynamics and linear gradient theory that was applied to derive a Knudsen number criterion and a new regime diagram (Dahms et al. 2016). The findings suggest that transition to a supercritical state arises as a result of thermal gradients within the interfacial region broadening of the interface and reduced molecular mean free path. The dynamics of the interfacial forces were also shown to gradually decrease as the interface broadens once it enters the continuum regime. This generalized framework provides a theoretical foundation to better understand and model the fuel injection interfacial transition dynamics in regions above the fluid critical condition. More recently, Poursadegh et al. (2017) conducted both experimental and theoretical analysis to show that under typical Diesel-engine relevant conditions, the time scale for heating up the jet by surroundings is much shorter than the jet breakup time scale, which explains the turbulent mixing behaviour observed in experiments. To provide insights into the high-pressure combustion systems, accurate and robust simulation tools are required. The transcritical nature of flows in a diesel injection process has motivated recent studies to utilize the diffuse-interface method.

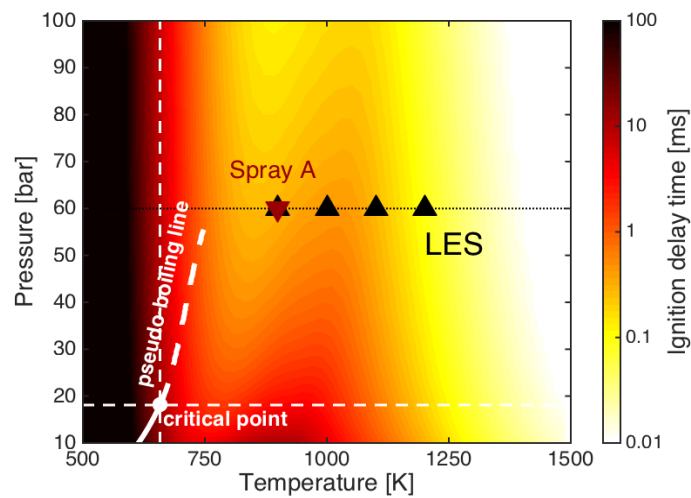

**Fig. 1.** Ambient operating conditions considered in this study (labeled by triangles). Critical point and pseudo-boiling line for *n*-dodecane are shown as white dot and lines, respectively. Ignition delay time for *n*-dodecane/air mixture at $\phi$ = 1.0 from homogeneous reactor simulations are shown as contours

In contrast to sharp interface techniques, where interfaces are explicitly tracked or resolved in the computational domain, this method artificially diffuses the interfaces. This is attractive for transcritical flows where interfaces are not present. However, it remains an open research question whether interfacial flows or droplets exist under conditions relevant to real applications (Yang 2000; Banuti et al. 2017). Oefelein et al. (2014) presented a large-eddy simulation (LES) framework coupled with real-fluid thermodynamics and transport to show that for typical diesel conditions the dynamics of the dense liquid jet mixing layer is dominated by real-fluid effects. Their findings showed that in case of typical diesel conditions the mixing path associated with all states during injection does not cross the liquid-vapor regime, suggesting that interfacial mixing layer dynamics are locally supercritical. Using similar numerical techniques, Lacaze et al. (2015) further analyzed the details of the transient mixing and processes leading to auto-ignition during diesel injection. It was suggested that the large density ratio between the supercritical fuel and the ambient gas leads to significant penetration of the jet with enhanced turbulent mixing at the tip and strong entrainment effects. Also reported was the presence of supersonic regions in the mixing layer due to the significant decrease in the speed of sound due to the real-gas thermodynamics, turbulent mixing, and high injection velocity. Knudsen et al. (2017) employed a compressible Eulerian model to describe the liquid fuel injection process under consideration of the internal nozzle flow and compressibility effects. Matheis and Hickel (2017) developed a thermodynamic model where phase separation is considered through vapor-liquid equilibrium calculations in a fully-conservative diffuse-interface method.

For reacting sprays, Lagrangian droplet models have been used in conjunction with a gas-phase combustion model, for which flamelet-based models and transported probability density function (TPDF) methods are typically employed. Simulations of multiple-injection realizations were conducted by Pei et al. (2015) using the TPDF method, showing that the first ignition was initiated in a lean mixture and subsequently propagated to the rich mixture. Using LES, coupled with a Flamelet Generated Manifold combustion model, Wehrfritz et al. (2016) investigated the early flame development with respect to temperature and formaldehyde at near-Spray A conditions. They reported that auto-ignition taking place in multiple stages denoted as first, second, and flame development stage, and observed formaldehyde formation prior to ignition at the tip of the fuel-rich gas jet consistent with experiments. More recently, a coupled LES and Tabulated Flamelet Model with multiple realizations was utilized to study the flame structure and ignition dynamics at low-temperature combustion (LTC) conditions in the temperature range between 750-1100 K (Kundu et al. 2017). Significant differences in flame structure were found at the low-temperature condition (750 K), including the formation of formaldehyde in lean regions from first-stage ignition. At typical engine conditions (900 K), formaldehyde formation was observed in the rich region followed by OH and high temperature in the stoichiometric region with higher scalar dissipation rates. Utilizing 1D unsteady flamelet calculations with detailed chemistry, Dahms et al. (2017) provided a conceptual model to describe the turbulent ignition process in high-pressure spray flames, and demonstrated the significance of turbulence-chemistry interactions under these conditions.

The objective of this study is to combine the diffuse-interface method with finite-rate chemistry model to investigate diesel-fuel injection and auto-ignition processes under realistic engine conditions. The experimental setup and computational domain are described in Sec. 2. The numerical setup is presented in Sec. 3. Governing equations, thermo-transport models, and the chemical mechanism employed in this work are discussed, which followed by a discussion of the numerical methods and boundary conditions. In Sec. 4, simulation results are presented and compared with available experimental data to examine the performance of the developed numerical framework. The paper finishes with conclusions in Sec. 5.

## 2. Case description

### 2.1 Spray A configuration

In this study, the Spray A single hole injector configuration is considered, which is the target case of the Engine Combustion Network (ECN) (Pickett et al. 2011). The experimental apparatus of the spray combustion chamber at Sandia National Laboratories is shown in Fig. 2. The single hole diesel injector has a nominal diameter of 90 µm and is operated with pure *n*-dodecane fuel at a rail pressure of $p_{rail}$ = 1500 bar.

The ambient operating conditions considered in this study are illustrated in the *p-T* diagram as shown in Fig. 1. Both non-reacting and reacting cases are considered, corresponding to 0% and 15% ambient $O_2$ compositions, respectively. Liquid *n*-dodecane fuel is injected at a subcritical temperature of $T_{inj}$ = 363 K into a hot ambient environment at a pressure of $p_\infty$ = 60 bar. An ambient temperature of $T_\infty$ = 900 K is considered for the non-reacting case (Spray A conditions) and effects of LTC on the auto-ignition are examined by performing simulations for a series of ambient temperatures varying from 900 K to 1200 K. At these conditions, the liquid *n*-dodecane undergoes a transcritical injection process before auto-ignition, during which the liquid fuel is heated and mixed with the ambient gaseous environment.

### 2.2 Computational domain

A three-dimensional cylindrical computational domain with a diameter of 40 mm and a length of 80 mm is used in this study. The injector geometry is not included in the computational domain and boundary conditions are provided at the exit of the injector nozzle. A structured mesh with hexahedral elements is used. The mesh is clustered in the region near the injector along the shear layers, and stretched in downstream and radial directions. The minimum grid spacing is 4 µm near the nozzle exit, which results in approximately 20 grid points across the injector nozzle. The maximum grid spacing downstream is less than 0.05 mm. The current grid resolution at locations where auto-ignition occurs is able to resolve the ignition kernel length scale estimated by non-premixed flame calculations. The total cell count of the mesh is 8.7 millions.

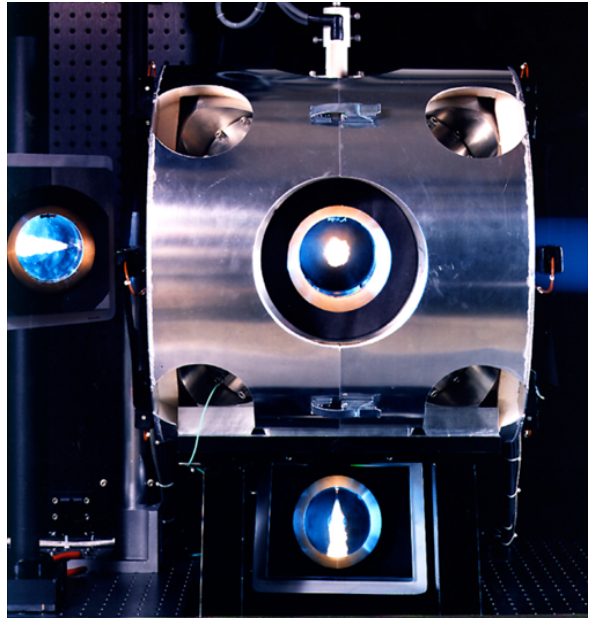

**Fig. 2.** Sandia Spray Combustion Chamber (Picket et al. 2011)

## 3. Numerical setup

### 3.1 Governing equations

The governing equations for the diffuse-interface method are the Favre-filtered conservation laws for mass, momentum, total energy, and species, taking the following form

$$\frac{\partial \bar{\rho}}{\partial t} + \nabla \cdot (\bar{\rho}\widetilde{\boldsymbol{u}}) = 0 \,, \tag{1a}$$

$$\frac{\partial \bar{\rho}\widetilde{\boldsymbol{u}}}{\partial t} + \nabla \cdot (\bar{\rho}\widetilde{\boldsymbol{u}}\widetilde{\boldsymbol{u}} + \bar{p}\boldsymbol{I}) = \nabla \cdot \bar{\boldsymbol{\tau}}_{v+t} \,, \tag{1b}$$

$$\frac{\partial \bar{\rho}\widetilde{E}}{\partial t} + \nabla \cdot \left[\widetilde{\boldsymbol{u}}(\bar{\rho}\widetilde{E} + \bar{p})\right] = \nabla \cdot (\bar{\boldsymbol{\tau}}_{v+t} \cdot \widetilde{\boldsymbol{u}}) - \nabla \cdot \bar{\boldsymbol{q}}_{v+t} \,, \tag{1c}$$

$$\frac{\partial \bar{\rho}\widetilde{Y}_k}{\partial t} + \nabla \cdot (\bar{\rho}\widetilde{\boldsymbol{u}}\widetilde{Y}_k) = -\nabla \cdot \bar{\boldsymbol{J}}_{k,v+t} + \bar{\dot{\omega}}_k \,, \tag{1d}$$

where $\rho$ is the density, $\boldsymbol{u}$ is the velocity vector, $p$ is the pressure, $E$ is the specific total energy, $\boldsymbol{\tau}$ is the stress tensor, $\boldsymbol{q}$ is the heat flux, and $Y_k$, $\boldsymbol{J}_k$, and $\dot{\omega}_k$ are the mass fraction, diffusion flux, and chemical source term for species $k$, and the species equations are solved for $k = 1, \cdots, N_S - 1$ where $N_S$ is the number of species. Subscripts $v$ and $t$ denote viscous and turbulent quantities, respectively. The system is closed with a state equation, $\bar{p} = f(\bar{\rho}, \widetilde{T}, \widetilde{Y}_k)$, where $T$ is the temperature. The subgrid-scale effects in the equation of state are neglected in the current study.

### 3.2 Thermodynamic relations

For computational efficiency and for accurately representing properties near the critical point (Miller et al. 2001), the Peng-Robinson cubic state equation (Peng and Robinson 1976) is used in this study, taking the form

$$p = \frac{RT}{v - b} - \frac{a}{v^2 + 2bv - b^2}, \tag{2}$$

where $R$ is the universal gas constant, $v$ is the specific volume, and the coefficients $a$ and $b$ are dependent on temperature and composition to account effects of intermolecular forces, and are evaluated as

$$a = \sum_{i=1}^{N}\sum_{j=1}^{N} X_i X_j a_{ij}, \tag{3a}$$

$$b = \sum_{i=1}^{N} X_i b_i, \tag{3b}$$

where $N$ is the number of species and $X_i$ is the mole fraction of species $i$. Extended corresponding states principle and pure fluid assumption for mixtures are adopted (Ely and Hanley 1981, 1983). Parameters $a_{ij}$ and $b_i$ are evaluated using the recommended mixing rules by Harstad et al. (1997). Procedures for evaluating thermodynamic quantities such as internal energy, specific heat capacity and partial enthalpy using the Peng-Robinson state equation are described in detail in Ma et al. (2014, 2017).

The dynamic viscosity and thermal conductivity are evaluated using Chung's method with high-pressure correction (Chung et al. 1984, 1988). This method is known to produce oscillations in viscosity for multi-species mixtures when both positive and negative acentric factors are present for individual species (Hickey et al. 2013). To solve this problem, a mole-fraction-averaged viscosity that is evaluated from the viscosity of each individual species can be used. Takahashi's high-pressure correction (Takahashi 1974) is used to evaluate binary diffusion coefficients.

### 3.3 Chemical mechanism

The computational cost for the finite-rate chemistry model is directly related to the number of species and reactions included in the chemical mechanism of choice. Therefore, for large-scale turbulent simulations, the reaction chemistry requires dimensional reduction such that the number of transported scalars is maintained at a reasonable level.

In the present study, a 33-species reduced mechanism for *n*-dodecane/air combustion is used. This mechanism is reduced from a 54-species skeletal mechanism (Yao 2017). Zero-dimensional auto-ignition computations are performed at a pressure of $p$ = 60 bar, initial temperatures in the range of $T$ = 800-1000 K, and equivalence ratio in the range of $\phi$ = 0.5-2. The results are sampled to apply the level of importance criterion (Lovas 2000) using the YARC reduction tool (Pepiot et al. 2008). From this, 21 species are identified to be suitable for quasi-steady-state (QSS) approximation. A validation is provided in Fig. 3, showing that the prediction of ignition time is very close to the original skeletal mechanism for the whole range of temperature and equivalence ratio.

The reduced mechanism is incorporated into the CFD solver using a combustion chemistry library based on Cantera (Goodwin et al. 2011). The library allows for the run-time specification of QSS species and linearized quasi-steady state approximation (L-QSSA) (Lu and Law 2006) is applied to this selection of species. The sparse linear system for L-QSSA is solved efficiently by separating the construction of the elimination tree from factorization via Eigen (Guennebaud et al. 2010).

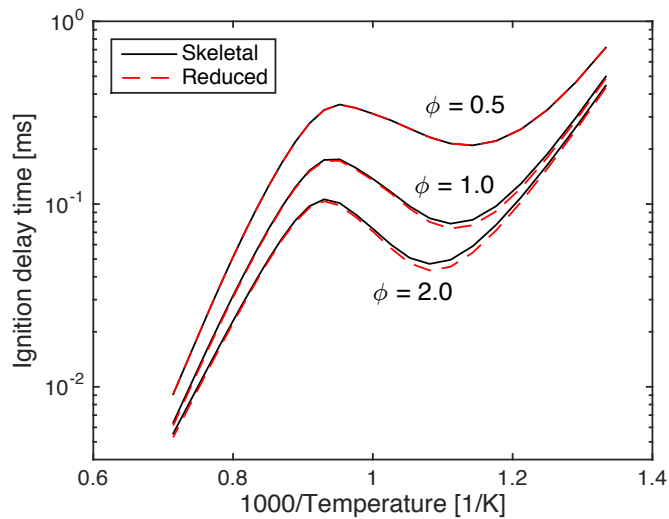

**Fig. 3.** Auto-ignition delay time of n-dodecane/air mixture predicted by 54-species skeletal mechanism (black solid line) and 33-species reduced mechanism (red dashed line) for three equivalence ratios at $p$ = 60 bar

**3.4 Numerical methods and boundary conditions**

The unstructured finite-volume solver, CharLES$^x$, is employed in this study. The convective fluxes are discretized using the sensor-based hybrid scheme with the entropy-stable flux correction technique developed in Ma et al. (2017). A central scheme, which is 4th-order accurate on uniform meshes, is used along with a 2nd-order ENO scheme. A density sensor is adopted. Due to strong non-linearities inherent in the real-fluid state equation, spurious pressure oscillations will be generated when a fully conservative scheme is used. To eliminate the spurious pressure oscillations, an adaptive double-flux method (Ma et al. 2017) is employed in this study. A second-order Strang-splitting scheme (Strang 1968) is applied to separate the convection, diffusion, and reaction operators. A strong stability preserving 3rd-order Runge-Kutta (SSP-RK3) scheme (Gottlieb 2001) is used for time integration of non-stiff operators. The reaction chemistry is integrated using a semi-implicit Rosenbrock-Krylov (ROK4E) scheme (Tranquilli 2014), which is 4th-order accurate in time and has linear cost with respect to the number of species. The stability of the ROK4E scheme is achieved through the approximation of the Jacobian matrix by its low-rank Krylov-subspace projection. As few as three right-hand-side evaluations are performed over four stages. Details about the development of the ROK4E scheme in the CharLES$^x$ solver can be found in Wu et al. (2018).

A Vreman sub-grid scale (SGS) model (Vreman 2004) is used as turbulence closure. The SGS turbulence/chemistry interaction is accounted for by using the dynamic thickened-flame model (Colin et al. 2000). The maximum thickening factor is set to be 4 in this study.

Fuel mass flux and temperature are prescribed at the injector nozzle exit using the time-dependent rate of injection as provided by the CMT virtual injection rate generator (Pickett et al. 2011), with default input parameters recommended by ECN for the Spray A case. A plug flow velocity profile is applied at the nozzle exit without synthetic turbulence. This choice of the inflow velocity profile was motivated by previous work of Lacaze et al. (2015) and effects of inflow turbulence on the injection dynamics are subject to future work. The pressure is prescribed at 60 bar at the outlet. Adiabatic boundary conditions are applied at all walls. All simulations are initialized with ambient conditions. A CFL number of unity is used during the simulation and a typical time step is about 0.6 ns. All simulations are conducted up to 1.2 ms after the injection.

# 4 Results and discussion

**4.1 Diesel-fuel transcritical injection at inert conditions**

The diesel-fuel transcritical injection process is studied through a non-reacting case with an ambient temperature of 900 K and simulation results are presented in this subsection. The liquid and vapor penetration lengths are extracted from LES using a threshold value of 0.6 and 0.01 for mixture fraction, respectively. Results up to 1 ms after injection are shown and compared with measurements in Fig. 4. The experimental vapor and liquid penetration lengths determined from Schlieren imaging and Mie scattering (Pickett et al. 2009; 2011) are also shown for comparison. It can be seen that the vapor penetration agrees with measurements favorably. It was found that the utilization of the inflow boundary conditions with a time-dependent fuel mass flux is essential for the accurate prediction. Another simulation with a constant mass flux without the initial ramp-up yielded appreciably longer penetration length, which is consistent with other studies (Pickett et al. 2011).

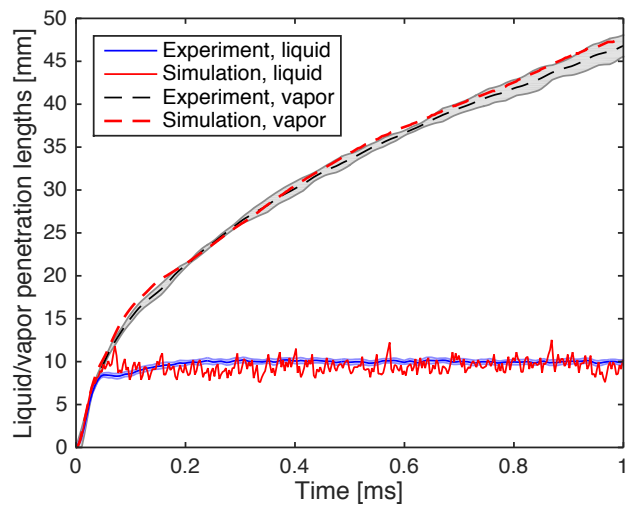

**Fig. 4.** Liquid and vapor penetration lengths predicted by LES in comparison with measurements (Picket et al. 2009) for the non-reacting case ($p_\infty$ = 60 bar, $T_\infty$ = 900 K)

For the validation of the liquid penetration length, a threshold value of 0.6 is used for the fuel mass fraction, providing good agreement with measurements. However, this threshold value is somewhat arbitrary. It was found that for mixture fraction values between 0.95 and 0.4, the predicted length varies between 5 mm and 15 mm. It was pointed out in the experimental investigations (Pickett et al. 2011; Manin et al. 2012) that the measured liquid penetration length is sensitive to measurement method, optical setup, threshold for determining liquid phase, and the actual geometry of the injector. Indeed, the resolution in either the simulation or the experiment is still not adequate to fully resolve the interfacial flow near the injector nozzle, if multi-phase flows do exist under these conditions.

The flow structures and mixing behaviors of the injection process further downstream are compared to the measurements of mixture fraction by Rayleigh scattering (Pickett et al. 2011). Multiple injections in the experiments provide ensemble-averaged statistics. In the simulation, the statistics of the steady period of injection are obtained by temporally averaging between 0.6 ms and 1.2 ms after the injection. Fig. 5 shows a comparison of the radial mixture fraction distribution at two different axial locations ($x$ = 17.85 and 25 mm). As can be seen in Fig. 5, there is a good agreement in the mean values of the mixture fraction at both locations, while the simulation predicts slightly higher rms values compared to the experimental data. These results along with the excellent agreement of the vapor penetration length are presented in Fig. 4, showing that the current numerical method is capable of predicting the turbulent mixing process between fuel and surrounding environment downstream of the injector after the dense liquid fuel is fully disintegrated.

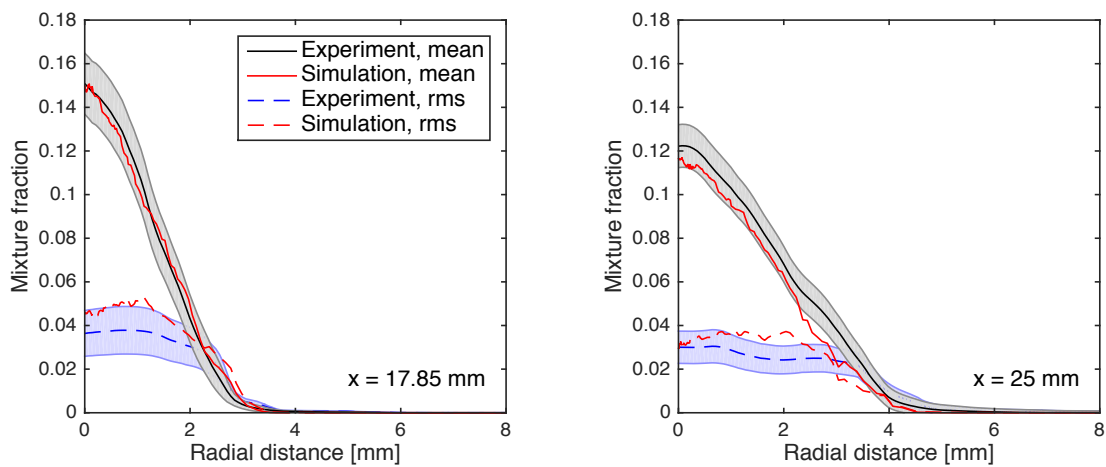

**Fig. 5**. Radial profiles of mean and rms values of mixture fraction in comparison with Rayleigh scattering measurements (Pickett et al. 2011) for the non-reacting case ($p_\infty$ = 60 bar, $T_\infty$ = 900 K)

### 4.2 Diesel-fuel transcritical autoignition

Following the evaluation of the solver for the non-reacting case, we now turn our attention to reacting cases. Simulation results for the case with an ambient temperature of 900 K (Spray A conditions) are shown in Fig. 6. Temperature fields are presented at several injection times along with the $CH_2O$ and OH fields. The results for $CH_2O$ and OH fields are plotted in the same figures to emphasize the spatial separation between them. It can be seen from the temperature results that the liquid *n*-dodecane fuel jet is heated up by the surrounding hot environment after being injected into the combustion chamber and a first-stage ignition can be observed at early ignition times (e.g., see temperature fields at 300 μs), which is associated with the rise in temperature and the formation of the $CH_2O$ species. After the second-stage ignition process (see results in Fig. 6 after 300 μs), high temperature regions with temperatures over 2000 K can be seen downstream the combustion chamber. From the species results in Fig. 6, it can be seen that $CH_2O$ is formed initially at the radial periphery of the jet. At later times, the maximum concentration of $CH_2O$ is observed in the center of the penetrating jet. The formation of OH is associated with the subsequent consumption of $CH_2O$ (Dahms et al. 2017) and the high-temperature chemistry (HTC) corresponding to the second-stage ignition process. High concentrations of OH are found near the edges of the penetrating jet due to the relatively low scalar dissipation rate and longer residence time in these regions (Kundu et al. 2017).

To assess the model performance in predicting intermediate species during the auto-ignition processes, Figs. 7 and 8 show results for mass fractions of $CH_2O$ and OH species along the centerplane at several injection times for the 900 K case in comparison with PLIF measurements (Maes et al. 2016). It can be seen from Fig. 7 that there is a good agreement between LES and measurements in terms of shape, magnitude, and location of the formation of $CH_2O$. Good agreement in OH fields can be observed between LES and experiments in Fig. 8, with a sharper representation in the simulation results.

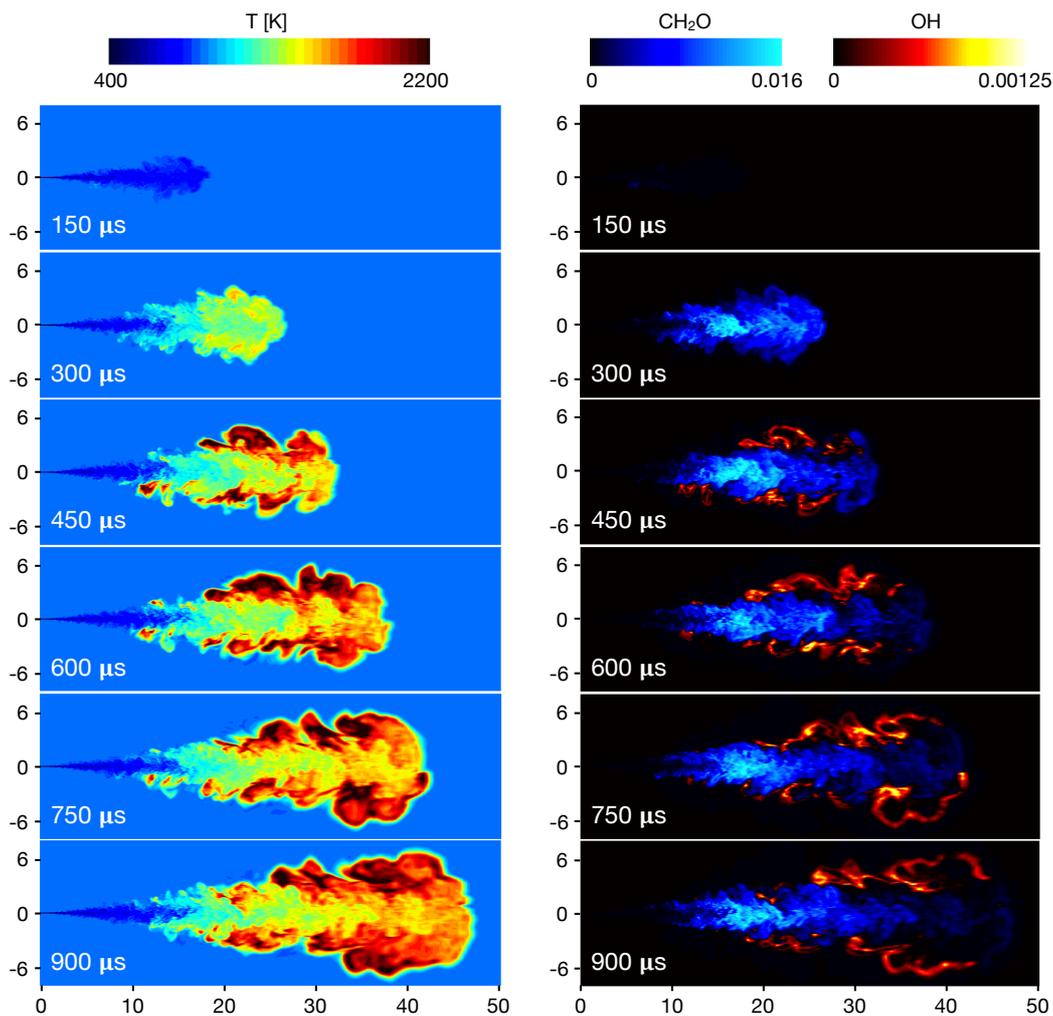

**Fig. 6.** Auto-ignition sequence of the case with 900 K ambient temperature (Spray A conditions) showing temperature and intermediate species fields. Spatial units in mm

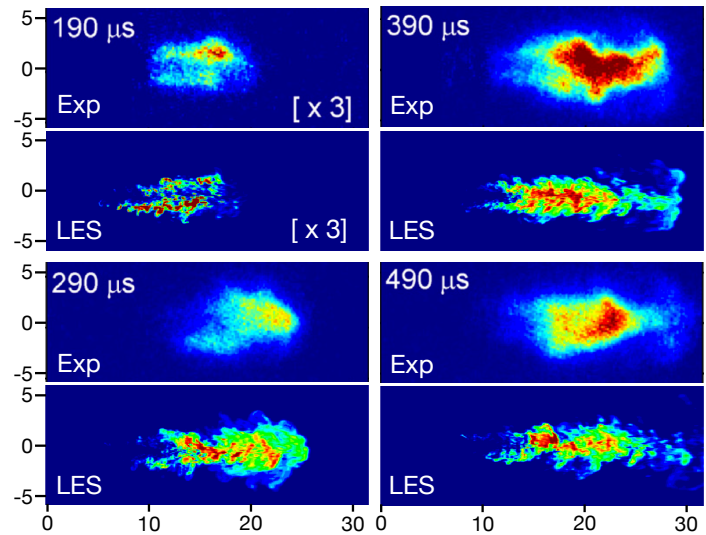

**Fig. 7.** Comparison between CH$_2$O mass fraction from LES results and false-color PLIF measurements (Skeen et al. 2015; Maes et al. 2016) at several injection times at 900 K ambient temperature (Spray A conditions). Spatial units in mm

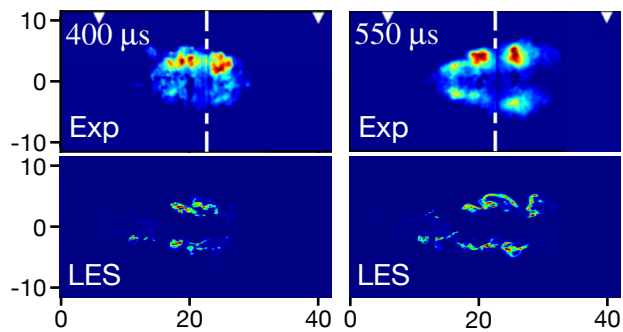

**Fig. 8.** Comparison of OH mass fraction from LES results and OH PLIF measurements (Pickett et al. 2011; Maes et al. 2016) at two injection times at 900 K ambient temperature (Spray A conditions). Spatial units in mm

Figure 9 shows the ignition delay time and the lift-off length predicted by LES at different ambient temperature conditions in comparison with measurements. Following the ECN-recommendations (Pickett et al. 2011) and criteria used by previous studies (Pei et al. 2015; Kundu et al. 2017; Bravo et al. 2016), the ignition delay time in LES is defined as the time when the maximum OH mass fraction reaches 14% of the value at quasi-steady state of the flame. The lift-off length is calculated using the line-of-sight OH mass fraction results from LES time-averaged between 0.8 to 1.2 ms during the quasi-steady state period, which is determined with a threshold of 14% of the level-off value similar to the procedures performed in the experiment (Pickett et al. 2011). In Fig. 9, the mean experimental data values from multiple experiments are shown with error bars indicating the range of the measurements. As can be seen from Fig. 9, good agreement is observed for both the ignition delay time and the lift-off length between LES and experiments, and the currently developed LES framework successfully captures the behaviors as a function of ambient temperature conditions. About 10% underprediction in ignition delay time from LES can be observed, which can be partly attributed to uncertainties in the chemical mechanism utilized. Note that shorter ignition delay times were also predicted by Yao et al. (2017) where the same parent skeletal chemical mechanism was adopted. Previous work utilizing flamelet-based combustion models (Wehrfritz et al. 2016) showed that the chemical mechanism has a significant effect on accurately predicting the ignition delay time.

Note that although the ignition delay time for the 1000 K and 1100 K cases are shorter than that for the 900 K case from homogeneous reactor calculations due to LTC (see Fig. 1), the actual ignition delay time for the three-dimensional injection process exhibits a monotonic behavior with respect to the ambient temperature, demonstrating the significance of turbulent mixing and heat transfer between fuel jet and the ambient prior to the ignition.

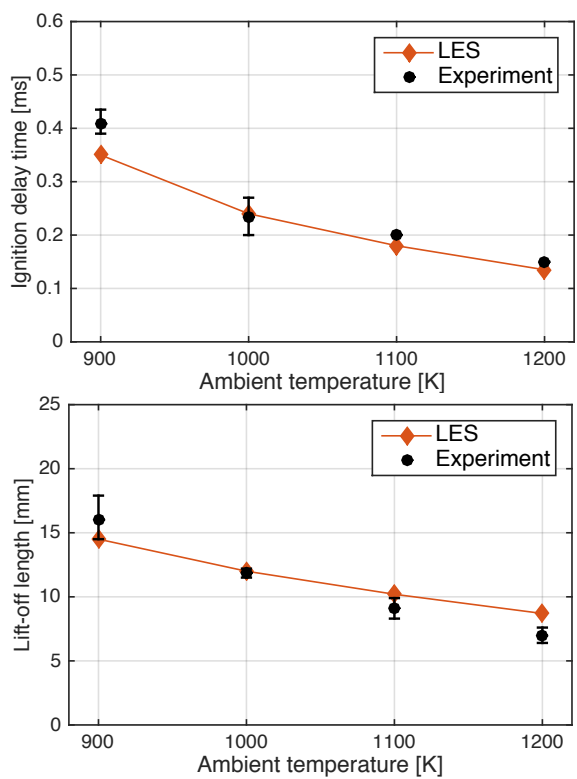

**Fig. 9.** Ignition delay time and lift-off length predicted by LES at different ambient temperature conditions in comparison with measurements (Pickett et al. 2011)

## Conclusions

In this work, a diffuse-interface method in conjunction with a finite-rate chemistry model is presented for the modeling of diesel fuel injection and auto-ignition processes under transcritical conditions. Compressible multi-species conservation equations are solved with a Peng-Robinson state equation and real-fluid transport properties.

LES-calculations are performed to simulate the ECN Spray A target configuration (Pickett et al. 2011) for both inert and reacting conditions. Simulation results are analyzed and compared with available experimental measurements. For the non-reacting case, predicted vapor and liquid penetration lengths are in good agreement with experimental results. Four ambient temperature conditions are considered for reacting cases. Good agreement of the ignition delay time and the lift-off length is obtained from simulation results at different ambient temperature conditions and the formation of intermediate species is captured by the simulations, indicating that the presented numerical framework adequately reproduces the corresponding LTC and HTC ignition processes under high-pressure conditions that are relevant to realistic diesel fuel injection.

## Acknowledgments

Financial support by ARL under award #W911NF-16-2-0170 and NASA under award #NNX15AV04A is gratefully acknowledged. This work is supported in part by resources from DoD High Performance Computing Modernization Program (HPCMP) FRONTIER Award at ARL through project "Petascale High Fidelity Simulation of Atomization and Spray/Wall Interactions at Diesel Engine Conditions." Simulations were performed on DoD Supercomputing platform "Excalibur" Cray XC40.